\documentclass[fleqn,twoside,headings]{article}
\usepackage[headings]{espcrc2}

\readRCS
$Id: espcrc2.tex,v 1.2 2004/02/24 11:22:11 spepping Exp $
\usepackage{graphicx}
\usepackage[figuresright]{rotating}

\newcommand{\AmS}{{\protect\the\textfont2
  A\kern-.1667em\lower.5ex\hbox{M}\kern-.125emS}}

\title{Prospects for Upgrade of KEKB
\thanks{University of Cincinnati preprint \# UCHEP-08-07}
\thanks{Talk presented at the BEACH 2008 conference, Columbia, South Carolina 22-28 June, 2008} 
}

\author{K. Kinoshita\address[MCSD]{Department of Physics, University of Cincinnati \\ 
        P.O. 210011, Cincinnati, OH, USA}%
        \thanks{Supported by DOE grant \# DE-FG02-84ER40153.}}
       
\begin{document}

\begin{abstract}
The Belle experiment at the KEKB electron-positron collider is expected to have collected close to one billion $\Upsilon$(4S) events by the time it comes to an end in 2009.  
An upgrade to KEKB has been proposed.
It is designed for an order of magnitude higher luminosity than KEKB, following a three-year construction period.
The ultimate goal of $8 \times 10^{35}{\rm cm}^{-2}{\rm s}^{-1}$ luminosity would be reached through further improvements over several years.  
To exploit the physics accessible through this improved luminosity, an upgrade of the Belle detector is also planned.  
A new international collaboration, temporarily named sBelle, is in the process of being formed.  
Super-KEKB and sBelle were officially placed on the KEK 5-year Roadmap in early 2008.\vspace{1pc}
\end{abstract}

\maketitle

The richness of the physics in the Upsilon region in $e^+e^-$ collisions is well-known; the first Upsilon resonance was discovered in the late '70's and, by the time  the $B$-factories started up in 1999, hundreds of associated papers had been published.
The Upsilon mine has yet to be depleted;  for example, the Belle collaboration has submitted 265 articles for publication in peer-reviewed journals as of June 2008.
The most influential among these include many investigations of the $CKM$ matrix, including measurements of the $CP$-violating phase parameter $\sim 2\phi_1$ and constraints on $\phi_2$ and $\phi_3$, observation of new charmonia and charmonium-like states, discovery of $D^0$ mixing, and many probes for ``New Physics.''
The events, which are clean, numerous, and triggered on with high efficiency, contain $B$-mesons, many types of charm hadrons, tau leptons, 2-photon events, Upsilons ($\Upsilon$(4S), $\Upsilon$(10860), $\Upsilon$(3S), $\Upsilon$(1S)), and $B_s$ mesons.
Among the physics topics addressed through them are $CP$ violation and other weak physics, QCD, heavy quark spectroscopy, lepton flavor violation, and Dark Matter.

In 2009 the Belle experiment at KEKB will conclude, having collected approximately one billion $\Upsilon$(4S) events and achieved, even surpassed, its original scientific objectives. 
A proposed upgrade, with luminosity  two orders of magnitude greater than KEKB, is in the process of approval and would extend the heavy quark program by more than a decade, starting in 2012.

\section{Why Continue the Flavor Physics Program?}

While the $B$-factories have been highly successful, it is useful to ask what further insights could be achieved with up to two orders of magnitude more data.
Most straightforward are extensions of measurements and searches that are currently statistically limited.
Some of the more prominent examples are $B^0\to\rho^0\rho^0$ (which strongly constrains $\phi_2$), the Dalitz analyses used to obtain $\phi_3$, radiative $b\to s$ and $b\to d$ decays, and lepton-flavor-violating tau decays.
More intriguing, though, is the sensitivity of this region to new particles and forces.
The Standard Model (SM) rates for decays involving loop processes are generally suppressed due to CKM cancellation.
Furthermore, the $CP$ asymmetries of many accessible decay modes are expected to be small.
Previously unknown processes at higher mass scales and with complex couplings could thus contribute substantially to such decays and be revealed through precision measurements of rates and $CP$ asymmetries.
In the data accumulated to date, more than 1.4~ab$^{-1}$ from Belle and Babar combined, there are a number of deviations from expectation that hint at the possibility of New Physics.
By increasing the data two orders of magnitude, one can study rare decays and improve precision to the point of  probing physics well into the TeV mass scale.  

\section{Physics Goals for an Upgraded KEKB program}

KEK has proposed Super-KEKB, an upgrade of KEKB, with a goal of accumulating 50~ab$^{-1}$.
The LHC will commence operations later this year, and many significant measurements on $B$'s can be expected to come out of  LHC-b as well as the ATLAS and CMS experiments.
The primary advantages of the $B$-factory experiments for such measurements are the ability to detect photons and $K_L$'s and their so-called ``hermeticity,'' coverage of the detector over nearly $4\pi$ solid angle to contain the well-specified $e^+e^-$ annihilation event and thus enable identification of neutrinos.
The priority of the proposed physics program thus emphasizes measurements for which the $B$-factory has a clear advantage: $CP$ asymmetries in $b\to s$ and $b\to d$ penguin decays, lepton universality tests in decays such as $B^+\to \tau^+\nu$ and $B\to D^{(*)}\tau\nu$, and $CP$ asymmetries in $D$ mixing.


\begin{figure}[htb]
\begin{center}
\includegraphics[height=16mm]{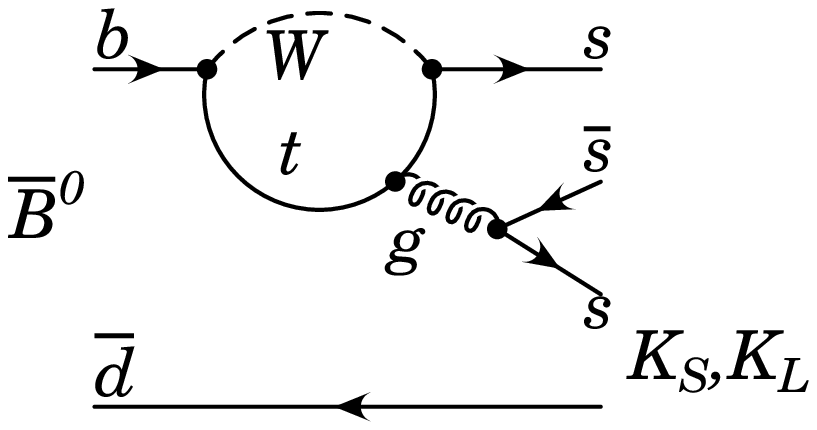}
\includegraphics[height=16mm]{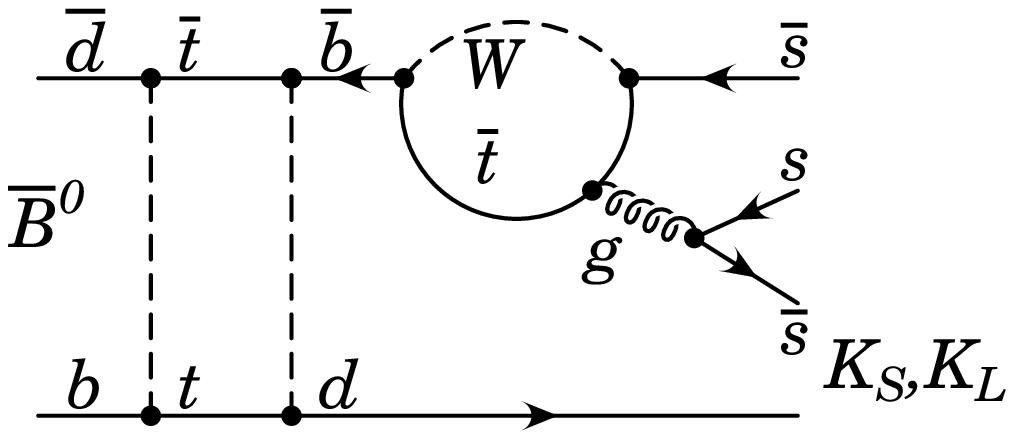}
\end{center}
\caption{Processes for decay $B\to s\bar s s \bar d$, (left) penguin and (right) mixing + penguin.}
\label{fig:Bmixing}
\end{figure}
\begin{figure}[htb]
\begin{center}
\includegraphics[width=70mm]{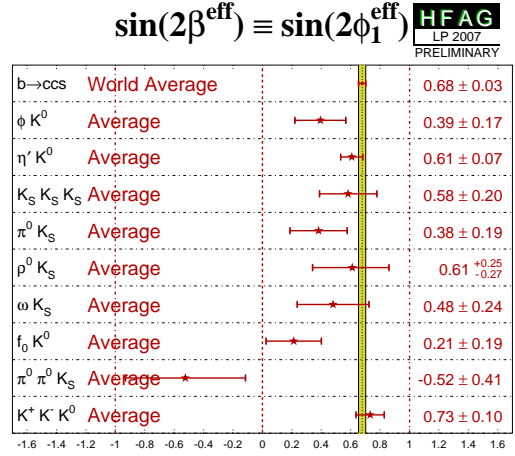}
\end{center}
\caption{HFAG compilation of $\sin 2\phi_1^{\rm eff}$ from $b\to s$ penguin decays.}
\label{fig:hfag}
\end{figure}
$B$ decays to a $CP$-eigenstate exhibit $CP$ asymmetry due to a non-zero complex phase angle between the amplitudes for the non-mixed and mixed processes.
For $b\to s\bar s s$, this is illustrated in Figure~\ref{fig:Bmixing}.
For the unmixed ``penguin'' process (left) $\bar B^0\to \{s\bar s\} K_{S/L}$, the dominant amplitude is proportional to the product $V_{tb}^*V_{ts}$, or $\approx -\lambda^2 A$ in the Wolfenstein parametrization\cite{Wolfenstein}.
The decay may also proceed through a mixing oscillation, $\bar B^0\to B^0 \{s\bar s\} K_{S/L}$ (Figure~\ref{fig:Bmixing}(right)), for which the amplitude is proportional to $V_{tb}^{*2}V_{td}^2 V_{tb}V_{ts}^*$.
The penguin parts of the two amplitudes differ by a particle conjugation, which in this case involves no relative complex phase.
The hadronic components of the two  are identical.  
The amplitude of mixing is well-measured and understood and includes the complex phase of $V_{tb}^{*2}V_{td}^2$ which,  expressed in terms of the angles of the Unitarity Triangle, is $2\phi_1$ (or $2\beta$).
The result is a $CP$-dependent decay rate
\begin{eqnarray}
{dN\over dt}(B\rightarrow f_{CP})\hspace{5cm}\nonumber\\
\ \ \ \propto{1\over 2}\Gamma e^{-\Gamma\Delta t}
[1+\eta_b\eta_{CP}{\rm sin}2\phi_1{\rm sin}(\Delta m\Delta t)],
\label{eqn:cpasym}
\end{eqnarray}
where $\eta_b=+1(-1)$ for a $B^0(\bar B^0)$,
$\eta_{CP}=+1(-1)$ if $CP$ is even (odd), $\Delta m$ is the mass difference between the two $CP$-eigenstates, and $\Delta t$ is the time 
interval from creation to the $CP$-eigenstate decay.
For the $b\to s\bar s s$ decays the penguin process involves three amplitudes, where the internal fermion in the loop is $u$, $c$, or $t$, and the unitarity of the CKM matrix would cause them to cancel were it not for the high mass of the $t$.  
The decay rate is thus suppressed, so that other forces with weaker but unsuppressed couplings could contribute to observable anomalous decay rates and $CP$ asymmetries.
The SM $CP$-asymmetry in particular is theoretically robust, so a precise measurement will have sensitivity to New Physics processes.

\begin{figure}[htb]
  \begin{center}
    \includegraphics[width=65mm]{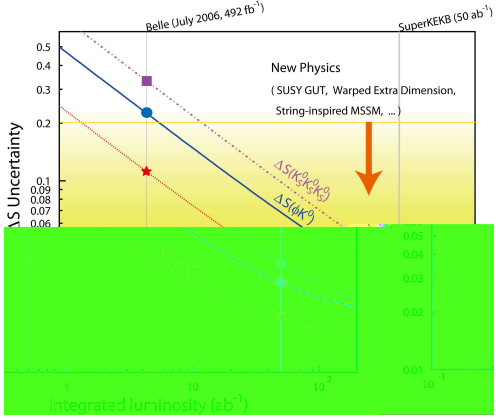}
  \end{center}
 \caption{Expected total errors on $\Delta{\cal S}$
   as a function of integrated luminosity.}
\label{fig:ds_da_sss_loi}
\end{figure}
The Heavy Flavor Averaging Group (HFAG) maintains a compilation of the world average asymmetry, measured as ``$\sin 2\phi_1^{\rm eff}$'' (or ``$\sin 2\beta^{\rm eff}$''), over all $b\to s$ hadronic penguin decays (Figure~\ref{fig:hfag}).
A na\"{i}ve averaging of the values gives $\sin 2\phi_1^{\rm eff}=0.56\pm 0.05$, which agrees to 2.2$\sigma$ or 3\% Confidence Level with the value measured in $b\to c\bar c s$ decays, $\sin 2\phi_1=0.680\pm 0.025$.
Possible reasons for the difference ($\Delta{\cal S}$) include statistical fluctuations or systematic errors, uncounted theoretical corrections, and New Physics.
The resolution of this possible inconsistency will require substantial additional data.
\begin{figure}[htb]
\begin{center}
  \includegraphics[width=35mm]{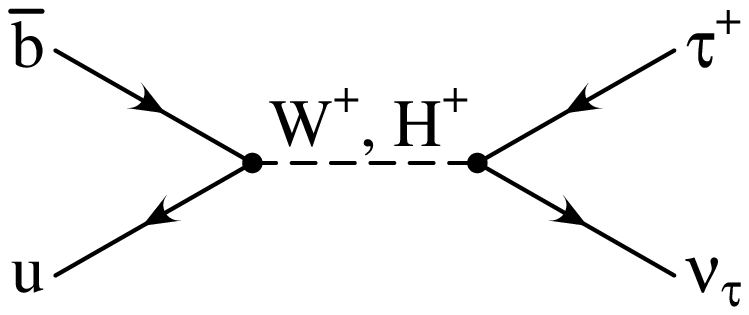}
  \includegraphics[width=28mm]{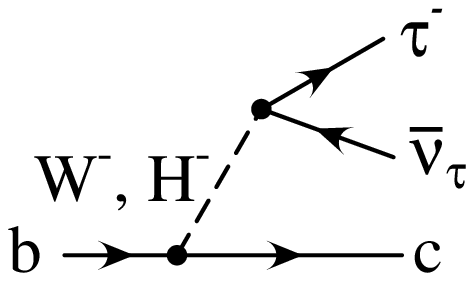} 
\caption{Processes for (left) $B\to\ell\nu$, (right) $B\to \{c\}\tau\nu$. 
}
\label{fig:taunuFD}
\end{center}   
\end{figure}
\begin{figure}[htb]
\begin{center}
  \includegraphics[width=49mm]{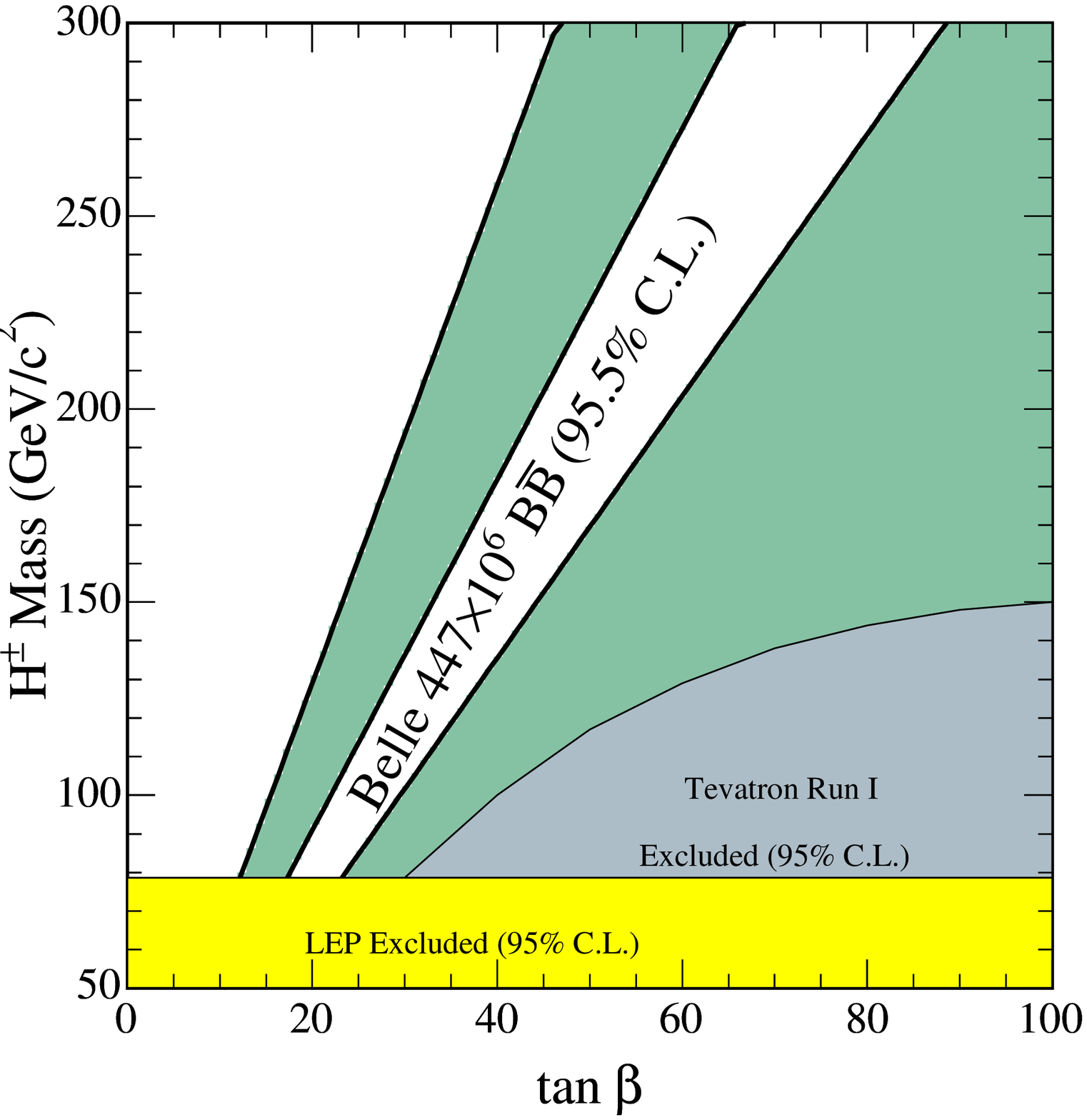}\\
  \includegraphics[width=44mm]{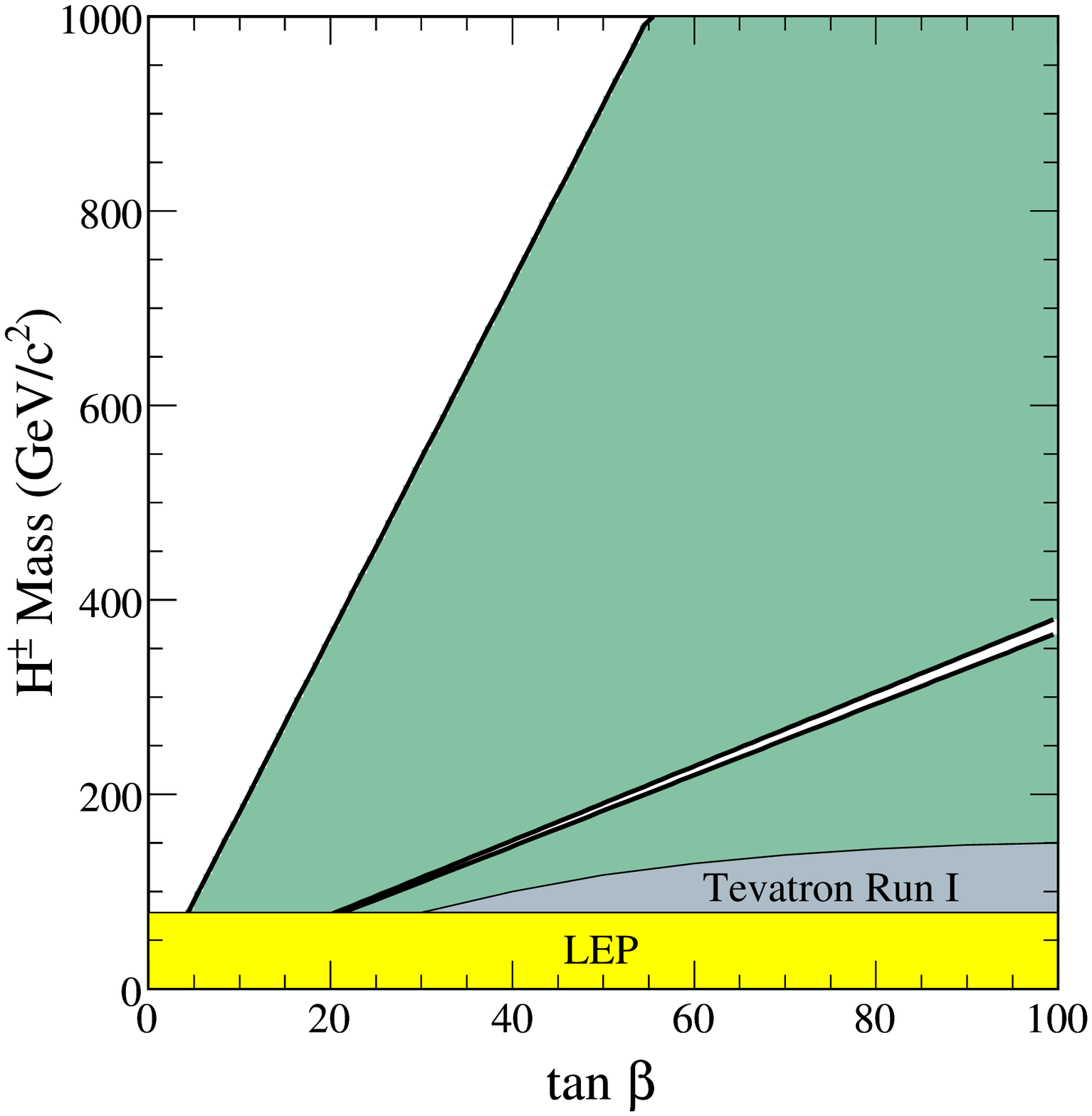} 
\caption{ (top) Region of $[M_{H^{+}}, \tan\beta]$ still allowed (white, 95.5\% CL) after Belle measurement of ${\cal B}(B^+\to \tau^+\nu_\tau)$ with 0.41~ab$^{-1}$.
(bottom) Allowed region if ${\cal B}(B^+\to \tau^+\nu_\tau)$ with 50~ab$^{-1}$ is consistent with the SM.  Note the change in vertical scale.
}
\label{fig:taunu}
\end{center}   
\end{figure}


Within the Standard Model the charged current coupling to leptons is universal.
For decays such as $B^+\to \ell^+\nu_\ell$ and $B\to D^{(*)}\ell\nu$ (Figure~\ref{fig:taunuFD}), a deviation of the ratio of the $\tau$ and $\mu$ modes from unity would thus indicate  New Physics.

For $B^+\to \tau^+\nu_\tau$ the SM partial width is theoretically robust, so that agreement between the measured and theoretical branching fractions  can be used to constrain the mass of the charged Higgs in some models.
The measured branching fraction for this channel\cite{taunu} is consistent with the SM.
In the two-Higgs-doublet model the rate is enhanced by a factor that depends on the Higgs mass and on $\tan\beta$, the ratio of vacuum expectation values of the two Higgs doublets\cite{wshou}.
In Figure~\ref{fig:taunu}(top) the region in $\tan\beta$ and Higgs mass that is not yet ruled out is shown in white. 
Much of this allowed region will be explored with 50~ab$^{-1}$ of $\Upsilon$(4S) data (Figure~\ref{fig:taunu}(bottom)).
The Standard Model predicts branching fractions of $1.6\times 10^{-4}$, $7.1\times 10^{-7}$, and $1.7\times 10^{-11}$ for $B^+\to \tau^+\bar\nu_\tau$, $B^+\to \mu^+\bar\nu_\mu$, and $B^+\to e^+\bar\nu_e$, respectively.
The  muonic decay will likely be observed with $\approx 5$~ab$^{-1}$ of data.

The decay $B\to D^{*}\tau\nu$ was recently observed at Belle, with ${\cal B}(B^0\to D^{*-}\tau\nu)=(2.0\pm 0.4\pm 0.4)\%$ \cite{Dtaunu}.
Its ratio with the corresponding muonic decay, which has been precisely measured for many years, is sensitive to the charged Higgs mass.
The $B^0\to D^{*-}\tau\nu$  and $B^+\to \tau^+\nu_\tau$  processes (Figure~\ref{fig:taunuFD}) have identical couplings to $\tau$ but differ on the quark side, so that a comparison of their behavior can also be used to distinguish between different New Physics scenarios, if a deviation is observed.

\begin{figure}[htb]
\begin{center}
\includegraphics[height=45mm]{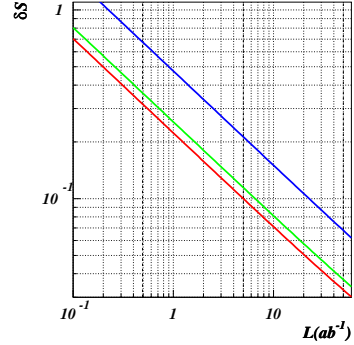}
\end{center}
\caption{Precision of $CP$-asymmetry as a function of integrated luminosity for decays $B^0\to K_S\pi^0\gamma$ (red), $B^0\to K^{*0}\gamma$
(green) and other $B^0\to K_S\pi^0\gamma$ (blue), extrapolated from Belle result on $B^0\to K_S\pi^0\gamma$\cite{btosgam_CP}, with 465~fb$^{-1}$.}
\label{fig:btosgam_CP}
\end{figure}

For a $b\to s\gamma$ decay with a $CP$-eigenstate in the final state hadronic component, the Standard Model final state is {\it not} a $CP$-eigenstate because the photon is left- (right-) handed for the $b\to s$ ($\bar b\to\bar s$) process.
The expected $CP$-asymmetry is thus much smaller than for gluonic penguin $b\to s$ modes, of the order of a few percent rather than $\sin 2\phi_1$.
Any sizable $CP$-asymmetry in such modes is thus a signature of right-handed currents in the $b\to s$ transition.
The expected precision in the asymmetry $S$ as a function of integrated luminosity, extrapolated from Belle's result for $B^0\to K_S\pi^0\gamma$\cite{btosgam_CP}, is shown in Figure~\ref{fig:btosgam_CP}.
  
\begin{figure}[htb]
\begin{center}
\includegraphics[height=50mm]{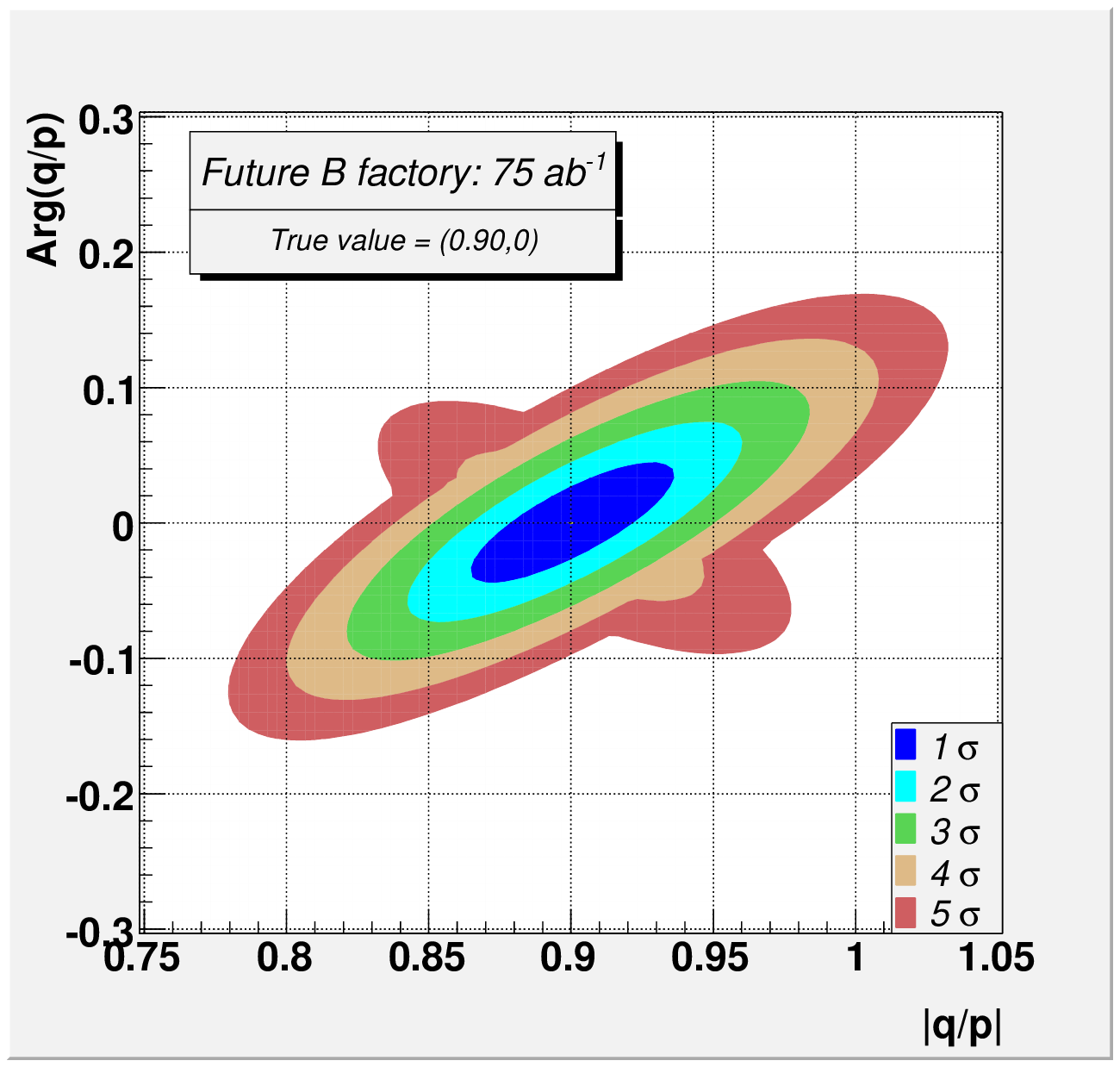}\\
\includegraphics[height=50mm]{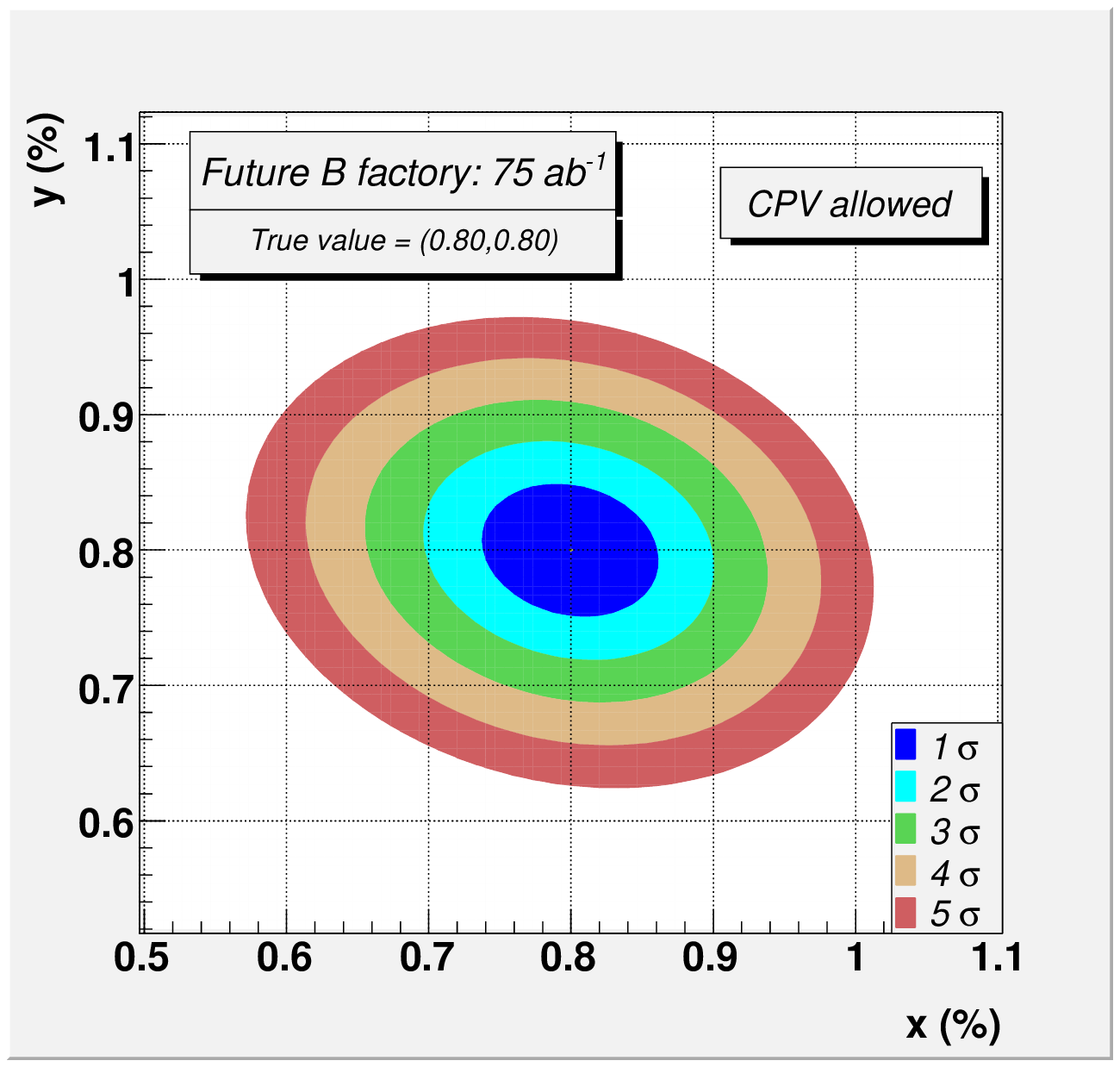}
\end{center}
\caption{Significance contours for (top) $D$ mixing parameters, $y\ vs.\ x$, with $x=0.8,\ y=0.8$, and (bottom) $CP$ violation parameters,  $Arg(q/p)\ vs.\ |q/p|$, with  $|q/p|=0.9,\ Arg(q/p)=0$, based on 75~ab$^{-1}$ of simulated data.
 }
\label{fig:Dmixing}
\end{figure}
The charm sector has long been acknowledged as a window to New Physics because of the strong suppression of loop rates and $CP$ asymmetry that is rooted in the unitarity of the CKM matrix.
With the recent observation of mixing in the $D^0$\cite{Dmixing}, it is clear that the $B$-factories have great sensitivity to rare charm processes and $CP$ asymmetries.  
The parameters pertaining to mixing are $x\equiv \Delta m/\Gamma$ and $y\equiv \Delta \Gamma/\Gamma$, where $\Delta m$ and $\Delta \Gamma$ are the differences between the masses and widths of the mass eigenstates, respectively, and $\Gamma$ is the average width.  
Shown in Figure~\ref{fig:Dmixing} are likelihood contours for projected measurements in $x-y$ and in the $CP$-violation parameters $q$ and $p$, based on simulations of 75~ab$^{-1}$.

\section{Design of a Detector for Super-KEKB}
With a goal toward  exploring beyond the horizon into the TeV mass range as well as leveraging current successes, a second-generation $B$-factory has been proposed at KEK.
The plan includes major upgrades of the KEKB collider and Belle detector.
The luminosity goals of ``SuperKEKB'' are ${\cal L}=8\times 10^{35}{\rm cm}^{-2}{\rm s}^{-1}$ instantaneous and $\int{\cal L}dt=50\ {\rm ab}^{-1}$ integrated luminosities,
to be achieved through multiple incremental improvements, each of which is designed to roughly double the luminosity yield.
They include improvements to the positron damping ring, superconducting RF, beam pipe design, crab cavities, and energy exchange C-band.
While it is hard to guarantee performance in a beam collider, the KEKB team has an excellent record of meeting and exceeding expectations.
In the current plan, which is presented as a baseline, the upgrade is phased so that it can be achieved with a continuation of the current level of funding.
Assuming a 3-year shutdown for construction starting in 2009 and 200 days per year of operations, it will be possible to collect 50ab$^{-1}$ by 2028.

A crucial step for KEKB is the implementation of ``crab'' cavities, to rotate the beams, which cross at a small angle, so that the beam overlap is maximized and collisions are ``head-on.''
Crab cavities were installed in the winter of 2007 and have been operating since  spring 2007.  
Thus far the principle of the crab design has been validated through the consistency of achieved specific luminosity with the predictions of simulations.
The maximum luminosity during the recent run has been $1.61\times 10^{34}{\rm cm}^{-2}{\rm s}^{-1}$, approximately equal to the maximum obtained without crab cavities.
To achieve the projected factor of two gain in luminosity, KEKB is working to improve the maximum stable beam current.

\begin{figure}[htb]
\begin{center}
\includegraphics[height=80mm]{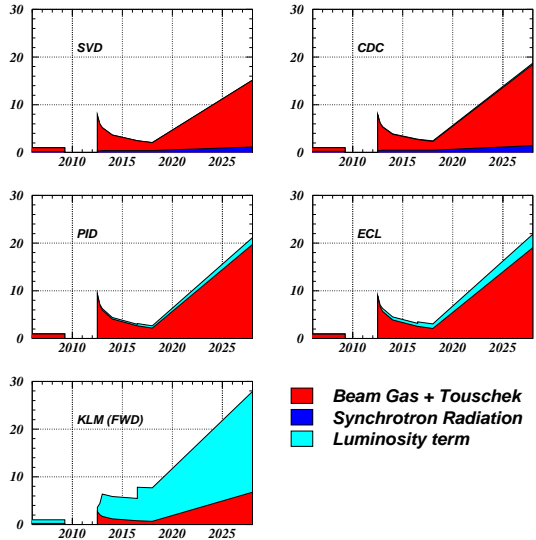}
\end{center}
\caption{Estimated background level as a function of year, normalized to total 2008 background levels in the Belle detector, based on proposed KEKB upgrade schedule. 
}
\label{fig:backgroundprojection}
\end{figure}

Increased beam current is accompanied by increases in background particles and events and synchrotron radiation, which in turn increase the rates to the detector for general radiation damage, detector occupancy, fake hits, pile-up, and triggers.
Figure~\ref{fig:backgroundprojection} shows the predicted rates for several types of background, based on projected currents and luminosities, in the major sectors of the Belle detector.
Most will experience a five-fold increase over current levels during early operations of the upgraded ring, increasing to approximately twenty-fold at maximum luminosity.

The upgrades to the Belle detector will focus on improving hadron identification and background tolerance.
An upgrade task force has been investigating these questions for several years. 
A Letter of Intent signed by nearly 300 physicsts from 61 institutions was submitted to KEK in 2004\cite{LOI}.  
KEK has launched a call for the formation of a new collaboration, temporarily named sBelle, to build and operate the upgraded detector.
The upgraded detector must be able to operate with a twenty-fold increase in background particles and a fifty-fold increase in real event rate over current conditions.
Modules targeted for upgrade include  data acquisition, silicon tracker, central drift chamber, particle ID device, calorimeter, and $K_L/\mu$ detector.
We describe briefly the current concept for each.

The silicon inner tracker is a critical component for vertexing but is at the same time exposed to high radiation doses and experiences high occupancy.
The upgrade design includes two additional outer layers which will have short strips to limit occupancy, fast electronics with reduced readout time (from 800~ns to 50~ns), and a pipelined readout.
As with silicon strip detectors in most current experiments, it is designed to be replaced periodically, and in a later version the inner layers will be closer to the beam line, 10~mm as opposed to the current 15~mm, and consist of pixel detectors, which are less vulnerable to radiation damage.

The central drift chamber will occupy a larger volume, having a larger inner radius (77~mm $\to$ 160~mm) to accommodate the new silicon tracker and an outer radius expanding into space previously occupied by aerogel and time-of-flight detectors (880~mm $\to$ 1140~mm).
To reduce occupancy each drift cell will be smaller, with an attendant increase in the number of sense wires to 15,140 from the current 8,400.

While the Aerogel threshold \v{C}erenkov  and time-of-flight counters have served Belle well, there is strong motivation to achieve better hadron separation for sBelle; for example, $K/\pi$ separation at momenta in the 2-4~GeV/$c$ range will need significant improvement to bring out a  $b\to d$ signal above $b\to s$.
To accomplish this, the current threshold \v{C}erenkov/TOF system in the barrel region will be replaced.
All of the designs under consideration involve the collection of \v{C}erenkov light from quartz glass radiators with sensitivity to the \v{C}erenkov angle.
The time-of-propagation (TOP) device measures the arrival time of photons, which depends on the time-of-flight from interaction point to the radiator and on the angle-dependent path length of the photon traveling the length of the bar via total internal reflection.
The ``imaging TOP'' concept adds a focusing mirror on one end and a photon collection array at the other that corrects for chromaticity and improves the performance.  
The ``focusing DIRC'' concept measures the angle more directly, through angle-dependent imaging of the light.
In addition to a new barrel detector, a ring-imaging Aerogel detector is planned for the endcap region.

While the current electromagnetic calorimeter, consisting of Tl-doped CsI crystals viewed by silicon photodiodes, achieves excellent resolution, modifications will be needed to maintain resolution at higher luminosities.
In the barrel the signal waveform will be sampled and fitted to avoid deterioration of resolution.
The relatively slow response ($\sim 1\mu$s) of Tl-doped CsI would result in degradation of signals in the endcap region due to pileup, the superposition of unrelated pulses over many beam crossings.  
The existing endcap will thus be replaced with pure CsI crystals, which have a rise time $\sim 30$ns, viewed by photomultipliers, which are well matched to the spectrum and speed of the CsI response.

The Belle $K_L/\mu$ detector, which consists of 5~cm steel plates interspersed with glass resistive plate chambers (RPC) running in streamer mode, has separate barrel and endcap systems.
The expected rates at the outer barrel region are sufficiently low that the existing chambers can be used at Super-KEKB if run in avalanche mode.  
The rates in the endcap regions will be beyond the capacity of the RPC's, so it has been proposed that the endcaps be replaced with plastic scintillator strips, with each layer in a 2-dimensional grid.

The ``KEK Roadmap'' outlines the long-range plan of the lab on a time scale of 10+ years.  
The Roadmap report released on January 4, 2008, by the director A.~Suzuki and the KEK management has included the KEKB luminosity and detector upgrades;
this report was affirmed by the external Roadmap Review Committe in March 2008.
A key feature in this preliminary version of the plan is that it accomplishes the upgrade with a flat funding profile.
According to the plan, KEKB and Belle will end datataking in 2009, and the activity will shift to Super-KEKB/detector construction for a three-year period, after which Super-KEKB operations will commence.  
The initial luminosity goal is $2\times 10^{35}{\rm cm}^{-2}{\rm s}^{-1}$.
Additional RF cavities will be installed as funding permits, raising luminosity over several years to $8\times 10^{35}{\rm cm}^{-2}{\rm s}^{-1}$.
While details are still under discussion between KEK and the Japanese funding agency (MEXT), the appearance of the KEKB upgrade as an official item in the Roadmap is highly significant.

The next step in the project will be to assemble an international collaboration to build and operate sBelle.  
Any institution is welcome to join and will have excellent opportunities to share in the construction and physics bounty.
Interested investigators are invited to contact anyone in the Interim Steering Committee.\footnote{Hiroaki Aihara (Tokyo/IPMU), Alex Bondar (BINP), Tom Browder (Hawaii), Paoti Chang (NTU), Toru Iijima (Nagoya), Peter Krizan (Chair, Ljubljana), Thomas Muller (Karlsruhe), Henryk Palka (Crakow), Christoph Schwanda (Vienna), Martin Sevior (Melbourne), Eunil Won (Korea), Changzheng Yuan(IHEP, China), Yutaka Ushiroda, Yoshi Sakai(KEK),  Masa Yamauchi (KEK)
}

\section{Summary}

The $B$-factories will have collected during their lifetime, 1999-2009, over 1.4 billion $B$ pairs at the $\Upsilon$(4S) resonance.  
Based on these data, the CKM matrix has been firmly established as the origin of $CP$ asymmetry in the weak interaction.
It has been possible to make multiple independent measurements of parameters to test the unitarity of the CKM matrix, including the three phase angles $\phi_1$, $\phi_2$, and $\phi_3$, as well as $|V_{ub}|$, with a precision that would have been considered optimistic a decade ago.  
Some internal inconsistencies in the global fit of CKM, as well as mysteries such as unexpected differences in $CP$ asymmetry among $B\to K\pi$ modes, may hint at breaks in the Standard Model.
By increasing the data by two orders of magnitude, one can study rare decays and improve precision to the point of  probing physics into the TeV mass scale.  
The KEKB upgrade, which will ultimately run at $8\times 10^{35}{\rm cm}^{-2}{\rm s}^{-1}$ luminosity will focus on $CP$-asymmetries in $B$ penguin decays, tests of lepton universality, lepton flavor violation, and precision measurement of all aspects of the CKM matrix.
The project has been included in the KEK Roadmap in 2008, a major step in bringing it to reality.
The proposed detector upgrade, currently named sBelle, has been under study for several years.  
The technical design will be finalized in the next year, and a new international collaboration is being formed.

\end{document}